\pdfoutput=1
\documentclass[aps,prl,10pt,twocolumn,notitlepage,superscriptaddress,footnoteinbib]{revtex4-2}
\usepackage[utf8]{inputenc}
\usepackage{amsmath,amssymb,amsfonts,graphicx,times,mathtools,amsthm,centernot}
\usepackage{lmodern}
\usepackage[a4paper,margin=2.4cm]{geometry}
\usepackage{amsmath,amssymb,amsthm,mathtools}
\usepackage{bm}
\usepackage{braket}
\usepackage{physics}
\usepackage{graphicx}
\usepackage[dvipsnames]{xcolor}
\usepackage{microtype}
\newcommand{\id}{\mathbbm{1}}

\usepackage[breaklinks=true,colorlinks=true,linkcolor=teal,urlcolor=teal,citecolor=teal]{hyperref}
\usepackage{orcidlink}

\usepackage{bm}
\usepackage{bbm}

\begin{document}
% ---------- Title ----------
\title{Precision limits for time-dependent quantum metrology under Markovian noise}
\author{Luca Previdi\,\orcidlink{0009-0008-6320-2079}}
\email{lucaprevidi0202@gmail.com}
\affiliation{%
 Dipartimento di Fisica e Astronomia, Università di Bologna, I-40126 Bologna, Italy
}%
\affiliation{Dipartimento di Fisica, 
Universit\`a di Milano, I-20133 Milano, Italia}%
\author{Francesco Albarelli\,\orcidlink{0000-0001-5775-168X}}
\email{francesco.albarelli@gmail.com}
\affiliation{Università di Parma, Dipartimento di Scienze Matematiche, Fisiche e Informatiche, I-43124 Parma, Italy}
\affiliation{INFN—Sezione di Milano-Bicocca, Gruppo Collegato di Parma, I-43124 Parma, Italy}

% \date{\today}
\date{May 18, 2026} 
% manual date so that it is always correct on arXiv even if recompiled in the future

\begin{abstract}
We derive ultimate precision bounds for estimating parameters encoded in \emph{time-dependent} Hamiltonians in the presence of general Markovian noise, allowing for arbitrary adaptive protocols with fast controls and noiseless ancillas.
Extending the minimization-over-purifications framework to time-varying continuous channels, we obtain a differential upper bound on the achievable quantum Fisher information (QFI) that can be evaluated at all times via semidefinite programming.
For parameter-independent noise, we prove a universal long-time scaling law: if the coherent (noiseless) dynamics yields $Q_{\mathrm{coh}}(T)\sim T^{2k}$, then under Markovian noise the QFI scales at most as $Q(T)\sim T^{2k}$ in the DHNLS regime, whereas in the DHLS regime it is fundamentally limited to $Q(T)\sim T^{2k-1}$.
We illustrate these behaviors on paradigmatic driven-qubit sensors, exhibiting $T^{4}$ and $T^{3}$ scalings under dephasing and spontaneous emission, respectively.
Finally, we provide explicit continuous exact and approximate quantum error correction constructions---supplemented by spin-squeezed probes---that asymptotically saturate the bounds, establishing their tightness.
\end{abstract}

\maketitle

\textit{Introduction.}---Quantum metrology exploits non-classical resources, such as entanglement and superposition, to push the boundaries of measurement precision beyond classical limitations~\cite{Helstrom1976,Wootters1981,Braunstein1994,Braunstein1996,Paris2009,Demkowicz-Dobrzanski2015a}.
The extreme sensitivity of quantum systems to external perturbations, often a hurdle for quantum control, may become a powerful asset for estimating macroscopic parameters.
In standard parameter estimation schemes relying on $N$ uncorrelated probes or an interrogation time $T$, classical protocols are fundamentally bounded by the standard quantum limit (SQL), where precision scales as $1/\sqrt{N}$ or $1/\sqrt{T}$.
In contrast, fully quantum strategies can asymptotically achieve the Heisenberg limit (HL), improving the scaling to $1/N$ or $1/T$~\cite{Holland1993,Giovannetti2004c,Giovannetti2006,Giovannetti2011}, with far-reaching applications ranging from gravitational wave detection~\cite{ligo2011gravitational} and atomic clocks~\cite{Yang2025b} to magnetometry~\cite{Kominis2008,Barry2020}.

Despite these theoretical promises, realistic quantum sensors inevitably suffer from decoherence.
Under Markovian noise, governed by the Gorini-Kossakowski-Sudarshan-Lindblad (GKSL) master equation~\cite{Gorini1976,Lindblad1976}, the performance can be severely degraded, often forcing the ultimate precision back to the SQL~\cite{Escher2011,Demkowicz-Dobrzanski2012,Demkowicz-Dobrzanski2014}.
Recent systematic analyses of time-independent signals have shown that the achievable precision is fundamentally dictated by the algebraic relationship between the signal Hamiltonian and the noise jump operators~\cite{Sekatski2017,Demkowicz-Dobrzanski2017,Zhou2018,Wan2022,Kurdzialek2023a,Mann2025,Gorecki2026}.
Specifically, if a time-independent Hamiltonian lies within the span of the Lindblad operators, known as the Hamiltonian-in-Lindblad-span (HLS) condition, the ultimate precision is bounded by the SQL, which can be saturated via approximate quantum error correction (AQEC)~\cite{Zhou2020}. 
Conversely, if the Hamiltonian-not-in-Lindblad-span (HNLS) condition holds, the HL can be perfectly recovered using exact quantum error correction (QEC)~\cite{Zhou2018}.

However, the metrological limits of \textit{time-dependent} signals under Markovian noise remain largely unexplored.
In noiseless scenarios, dynamically modulating the parameter-encoding Hamiltonian, often supplemented by adaptive control, can yield exceptional precision scalings that surpass even the standard HL in time, achieving variances that scale as $1/T^k$ with $k > 1$~\cite{Pang2017,Yang2017,Gefen2017,Naghiloo2017,Gefen2019}.
Whether, and under what conditions, these exceptional scalings survive in the presence of realistic open-system dynamics represents a fundamental open question in quantum parameter estimation.
% \cite{pang_optimal_2017, yang2017quantum, gefen, naghiloo2017achieving, gefen2019overcoming}.
This is a timely question, since time-dependent and oscillatory signals arise naturally in AC magnetometry, spectroscopy, and driven many-body sensing settings~\cite{Schmitt2017,Meinel2022,Gribben2025}.
Further, recent theoretical studies have highlighted how oscillatory signals lead to metrological tradeoffs and optimality questions beyond the time-independent setting, especially in the broadband setting~\cite{Polloreno2023,Allen2025,Dey2025a,Polloreno2026}.

In this Letter, we establish the ultimate precision bounds for time-dependent quantum metrology under general Markovian noise and provide the explicit error-correction strategies required to saturate them.
By extending the minimization-over-purifications (MOP) framework~\cite{Fujiwara2008,Escher2011,Demkowicz-Dobrzanski2012,Demkowicz-Dobrzanski2017,Kurdzialek2023a}
%  \cite{kurdzialek2023using} 
to time-varying continuous channels, we derive rigorous upper bounds on the quantum Fisher information (QFI) that are valid for arbitrary adaptive control strategies, which can be efficiently evaluated at all times via a semidefinite program (SDP)~\cite{Das2025}. 
% \cite{das_universal_2025}.
Crucially, we demonstrate that when the derivative of a time-dependent Hamiltonian with respect to the parameter of interest does not lie in the Lindblad span, the exceptional QFI scaling of the noiseless dynamics is asymptotically preserved.
Conversely, if the derivative of such Hamiltonian lies in the Lindblad span, the noise penalty reduces the exponent of the optimal time scaling by exactly one, a bound that may still surpass standard coherent limits.
Finally, extending standard recovery maps to the time-dependent regime, we show that these generalized bounds are not merely mathematical constructs, but are asymptotically saturable utilizing continuous QEC, AQEC, and spin-squeezed states~\cite{Ulam-Orgikh2001}.
% \cite{ulam-orgikh_spin_2001}.

\begin{figure}[t]
    \centering
    \includegraphics[width=1\linewidth]{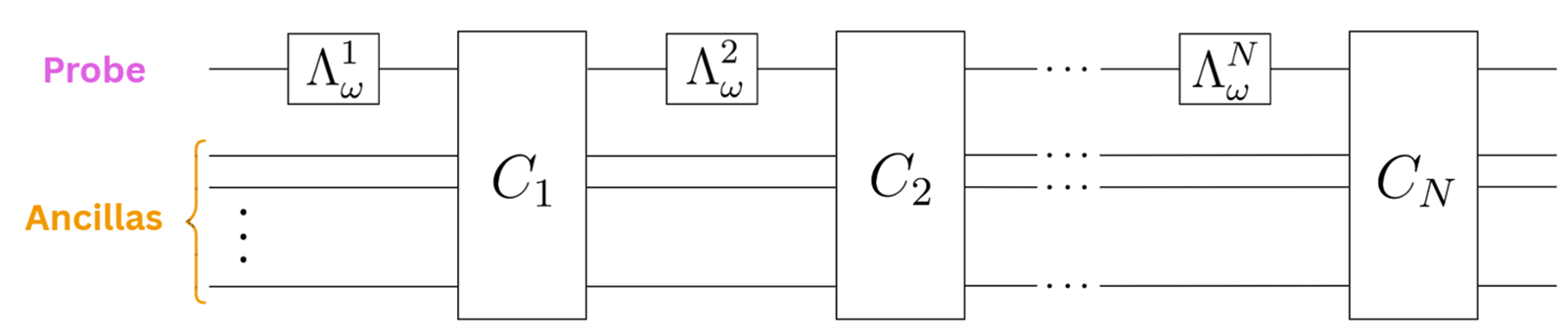}
    \caption{Pictorial representation of the most general adaptive strategy considered in this Letter.
    A probe evolves with a channel $\Lambda^i_\omega$ and then a unitary control is $C_i$ applied, coupling the probe to any number of noiseless ancillas.
    This step can be repeated an arbitrary number of times $N$.
    Notice that, differently from~\cite{Kurdzialek2023a}, the channels could be different each step, $\Lambda^i_\omega \neq \Lambda^j_\omega$, which is equivalent to consider a time-dependent channel.}
    \label{fig:adaptive_scheme}
\end{figure}

\textit{Precision bounds.}---We consider a time-dependent Markovian quantum channel, which depends on a parameter $\omega$, generated by a GKSL master equation. Denoting by $\rho_\omega(t)$ the state at time $t$, we assume
\begin{align}
 \notag\dot{\rho}_\omega(t)
= &
-i\bigl[H(\omega,t),\rho_\omega(t)\bigr]
\\& \notag+\sum_{j=1}^{J}\!\big(
L_j(\omega,t)\,\rho_\omega(t)\,L_j^\dagger(\omega,t)
\\&-\frac12\bigl\{L_j^\dagger(\omega,t)L_j(\omega,t),\rho_\omega(t)\bigr\}
\big),
\label{eq:lindblad}
\end{align}
where $\dot{\rho}_\omega(t)\equiv \partial_t\rho_\omega(t)$ and $\{\cdot,\cdot\}$ denotes the anticommutator.
To estimate $\omega$ we allow for a very general adaptive strategy (Fig.~\ref{fig:adaptive_scheme}): the probe undergoes the evolution~\eqref{eq:lindblad}, while we can intersperse the dynamics with an arbitrary number of unitary controls acting jointly on the probe and any number of noiseless ancillas.

In close analogy with the time-independent case~\cite{Demkowicz-Dobrzanski2014,Demkowicz-Dobrzanski2017,Kurdzialek2023a,Sekatski2017,Zhou2021,Kolodynski2013},
% \cite{demkowicz-dobrzanski_using_2014, demkowicz-dobrzanski_adaptive_2017, kurdzialek2023using, wan2022bounds, sekatski2016quantum, zhou_asymptotic_2021, fujiwara2008fibre, escher_general_2011, demkowicz2012elusive, kolodynski2013efficient}),
one can upper bound the instantaneous growth of the quantum Fisher information (QFI) $Q(t)$ achievable by such protocols as
\begin{equation}
\frac{d Q(t)}{dt}
\le
4\,\min_{\kappa(t),\,\eta(t),\,\gamma(t)}
\left(
\bigl\|\alpha(t)\bigr\|
+
\bigl\|\beta(t)\bigr\|\,\sqrt{Q(t)}
\right).
\label{eq:bound}
\end{equation}
Here $\|\cdot\|$ denotes the operator norm, and the minimization is over the auxiliary quantities $\kappa(t)$, $\eta(t)$ and $\gamma(t)$ introduced in the derivation.
We use the compact notation $ \mathbf{L}(\omega,t) \equiv [L_1(\omega,t),\ldots,L_J(\omega,t)]^{T}$ and the prime $(\cdot )'$ denotes the derivative with respect to $\omega$.
Moreover, we defined $ \alpha(t)
=
 [\eta(t)\,\mathbb{I}+\gamma(t)\,\mathbf{L}(\omega,t)+ i\,\mathbf{L}'(\omega,t) ]^{\!\dagger}
 [\eta(t)\,\mathbb{I}+\gamma(t)\,\mathbf{L}(\omega,t)+ i\,\mathbf{L}'(\omega,t) ]$ and $
-i\, \beta(t) =
H'(\omega,t)
-\frac{i}{2} (\mathbf{L}'^{\dagger}(\omega,t)\,\mathbf{L}(\omega,t)-\mathbf{L}^{\dagger}(\omega,t)\,\mathbf{L}'(\omega,t) )
+\kappa(t)\,\mathbb{I}
+\mathbf{L}^{\dagger}(\omega,t)\,\eta(t)+\eta^{\dagger}(t)\,\mathbf{L}(t)
+\mathbf{L}^{\dagger}(\omega,t)\,\gamma(t)\,\mathbf{L}(t)$ similarly to~\cite{Wan2022,Kurdzialek2023a}.
A proof of Eq.~\eqref{eq:bound} is presented in the Supplemental Material (SM)~\cite{supplementary}, but can alternatively be obtained from the results of~\cite{Wan2022}.
% \cite{wan2022bounds,kurdzialek2023using}.

We now specialize to the case of $\omega$-independent noise, i.e.\ $\mathbf{L}'(t)=0$.
Moreover, we assume that for $t\to\infty$ the Hamiltonian derivative admits the asymptotic expansion
\begin{equation}
H'(\omega,t)= t^{n}\,H_{0}(t)+O(t^{n-1}), 
\label{eq:Hprime_asympt}
\end{equation}
with $H_{0}(t)$ bounded for all $t$.
This assumption is discussed in more detail in the SM \cite{supplementary} and is valid for a wide class of Hamiltonians.
In this regime it is convenient to introduce $\alpha_{0}(t)
=
 [\eta(t)\,\mathbb{I}+\gamma(t)\,\mathbf{L}(t) ]^{\!\dagger}
 [\eta(t)\,\mathbb{I}+\gamma(t)\,\mathbf{L}(t) ],
-i\,\beta_{0}(t)=
t^{n}\,H_{0}(t)
+\kappa(t)\,\mathbb{I}
+\mathbf{L}^{\dagger}(\omega,t)\,\eta(t)+\eta^{\dagger}(t)\,\mathbf{L}(t)
+\mathbf{L}^{\dagger}(\omega,t)\,\gamma(t)\,\mathbf{L}(t).
$

We say that the system satisfies the derivative of the Hamiltonian in Lindblad span (DHLS) condition if there exist $T_{0}$ and a measurable set $I\subset (T_{0},\infty)$ whose complement has zero (Lebesgue) measure, such that for every $t\in I$ one can choose $\eta(t)$ and $\gamma(t)$ making $\beta_{0}(t)=0$.
If this is not possible, we say that the system is in the derivative of the Hamiltonian not in Lindblad span (DHNLS) regime.
Depending on which condition holds, the integration of the bound \eqref{eq:bound} admits different asymptotic forms as $T\to\infty$:
\begin{align}
 \notag& Q(T)\;\lesssim\;\mathcal{B}(T)  \\&\;
={\scriptsize \begin{cases}
\displaystyle
4\int_{0}^{T}
\min_{\substack{\kappa(t),\,\eta(t),\,\gamma(t)\\ \beta_{0}(t)=0}}
\bigl\|\alpha_{0}(t)\bigr\|\;dt
 \text{ for DHLS},\\[2.0ex]
\displaystyle
4\left[
\int_{0}^{T}
\min_{\kappa(t),\,\eta(t),\,\gamma(t)}
\bigl\|\beta_{0}(t)\bigr\|\;dt
\right]^{2}
 \text{for DHNLS}.
\end{cases}}
\label{eq:bound_DHLS_DHNLS}
\end{align}
Here $f(T)\lesssim g(T)$ for $T\to\infty$ means $\lim_{T\to\infty} f(T)/g(T)\le 1$.

Similarly to the time-independent scenario, the two regimes of Eq.~\eqref{eq:bound_DHLS_DHNLS} are intrinsically related to the well known concept of Lindblad span~\cite{Zhou2018,Zhou2020}:
\begin{align}
\mathcal{S}(t)
=\notag
\mathrm{span}_{\mathbb{R}}\!\big\{&
\mathbb{I},
\,L_j(t)^{\mathrm{H}},
\, i\,L_j(t)^{\mathrm{AH}},
\,\bigl(L_j^\dagger(t)L_k(t)\bigr)^{\mathrm{H}},\\&
\, i\,\bigl(L_j^\dagger(t)L_k(t)\bigr)^{\mathrm{AH}}
\big\},
\label{eq:lindblad_span}
\end{align}
where, for any operator $X$, $X^{\mathrm{H}}=(X+X^\dagger)/2$ is its Hermitian part and $X^{\mathrm{AH}}=(X-X^\dagger)/2$ is its anti-Hermitian part.
In particular, for fixed $t$ there exist $\kappa(t)$, $\eta(t)$ and $\gamma(t)$ such that $\beta(t)=0$ if and only if $H'(\omega,t)\in \mathcal{S}(t)$.

Finally, it is possible to prove a general asymptotic scaling statement (see SM \cite{supplementary}): for bounded-in-time jump operators, if in the coherent (noiseless) case the QFI scales as $Q_{\mathrm{coh}}(T)\sim T^{2k}$, then in the noisy case one has
\begin{equation}
Q_{\mathrm{DHNLS}}(T)\lesssim T^{2k}\,,
\qquad
Q_{\mathrm{DHLS}}(T)\lesssim T^{2k-1},
\end{equation}
which includes and generalizes the time-independent scenario~\cite{Huelga1997,Giovannetti2004c,Giovannetti2006,Giovannetti2011,Wan2022,Demkowicz-Dobrzanski2017,Kurdzialek2023a,Zhou2018,Zhou2020,Das2025} in the asymptotic long-time limit.
The DHNLS scaling will typically occur with a reduced prefactor, preserving the intuitive idea that a noiseless unitary evolution will be able to encode more information, under the (rather physical) assumption that the Lindblad operators do not depend on the parameter, at least not in a known way.
As we will show in the following sections both branches of Eq.~\eqref{eq:bound_DHLS_DHNLS} can be saturated in the large $T$ limit exploiting QEC strategies.

\textit{Representative models and scaling behavior.}---
As a concrete example, we consider two paradigmatic time-dependent signals widely employed in the literature~\cite{Gefen2017,Pang2017,Naghiloo2017,Hou2021a}:
% \cite{gefen, pang_optimal_2017,naghiloo2017achieving,hou_super-heisenberg_2021}: 
\begin{align}
    &H_{\mathrm{AC}}(\omega,t)=B\sin(\omega t)\sigma_z,\\
    &H_{\mathrm{RF}}(\omega,t)=B\bigl(\cos(\omega t)\sigma_x+\sin(\omega t)\sigma_z\bigr).
\end{align}
These represent, respectively, the Hamiltonian of a qubit coupled to an amplitude-modulated field and a qubit coupled to a rotating field.

As a first noise model, we consider a single jump operator of the form $ L=\sqrt{\varepsilon}\,\sigma_x,$
corresponding to dephasing along the $x$ direction. For the estimation of $\omega$, it is straightforward to verify that the DHNLS condition is satisfied for both $H_{\mathrm{AC}}$ and $H_{\mathrm{RF}}$. In this case, for $T\to\infty$ the QFIs read
\begin{equation}
\mathcal{B}_{\mathrm{AC}}=\mathcal{B}_{\mathrm{RF}}=\frac{4B^2}{\pi^2}T^4.
\end{equation}
Here, the maximum is taken over all possible strategies described in the previous section and pictorially represented in Fig. \ref{fig:adaptive_scheme}.
The QFI is the same for the two models because $ H_{\mathrm{RF},\perp}=H_{\mathrm{AC},\perp}=H_{\mathrm{AC}},$ and, as a consequence, it also coincides with the QFI obtained in the coherent case for $H_{\mathrm{AC}}$.
This shows that the $T^4$ scaling of the QFI observed in the noiseless scenario~\cite{Pang2017,Gefen2017} can be retained even in the presence of noise.

\begin{figure}
    \centering
    \includegraphics[width=1\linewidth]{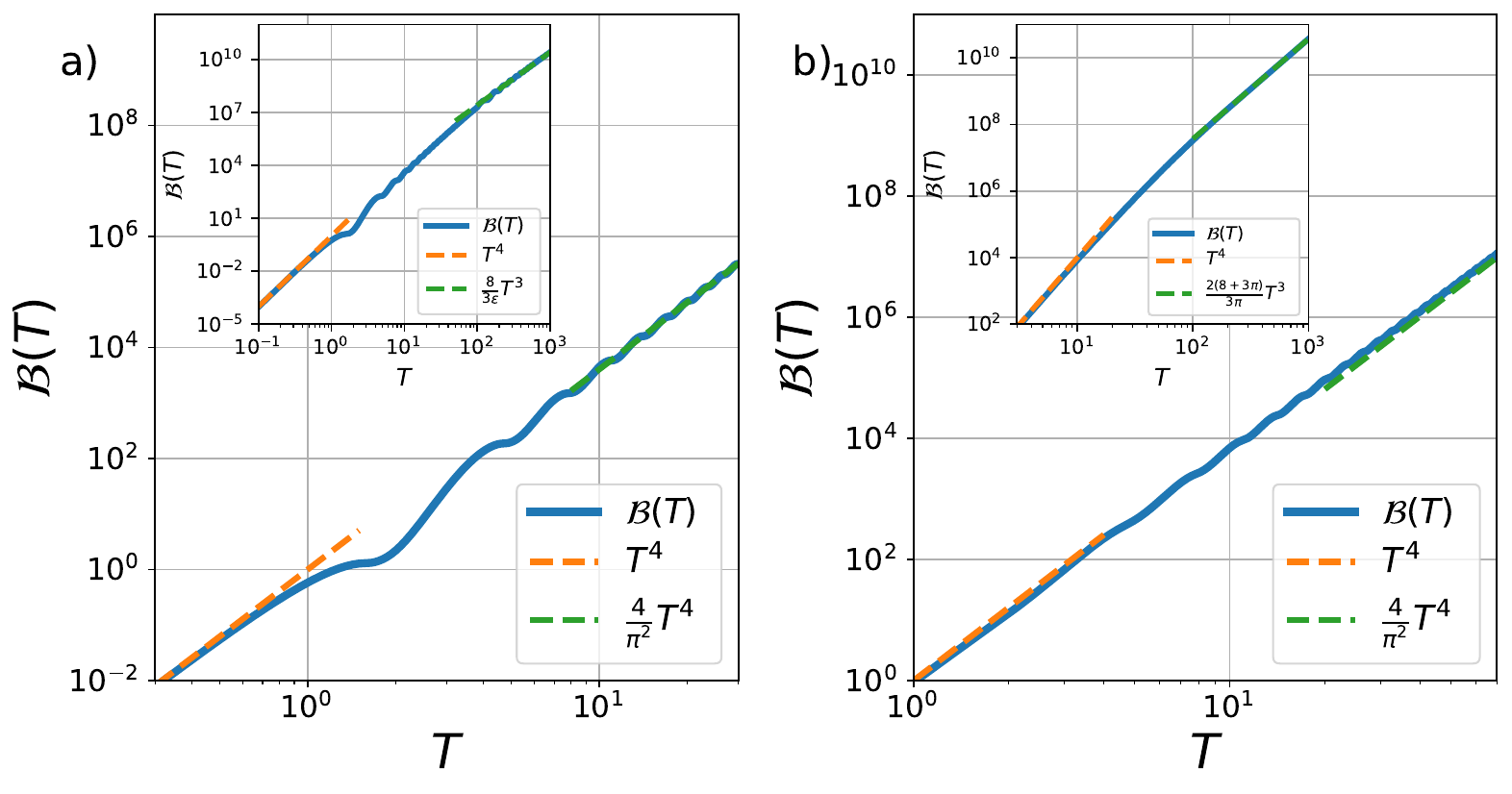}
    \caption{Plot of the full numerically optimized bound (solid blue) and its asymptotic behaviors (dashed lines) for $B = 1$, $\omega = 1$, $\varepsilon = 0.1$. Panel a) shows $H_\text{AC}$ under dephasing along the $x$ direction, while the inset depicts spontaneous emission noise.  Panel b) shows $H_\text{RF}$ under dephasing along the $x$ direction, while the inset depicts spontaneous emission noise. }
    \label{fig:sdp_bounds}
\end{figure}

A full numerical evaluation of the bound can be performed using the SDP (see SM \cite{supplementary}), and the result is shown in Fig.~\ref{fig:sdp_bounds}. One can observe an initial transient regime, after which the QFI crosses over to its large-time asymptotic behavior. While the prefactor is reduced, the QFI firmly maintains its $T^4$ scaling across both regimes.

As a second noise model, we consider a Lindblad operator of the form $ L=\sqrt{\varepsilon}\,\sigma_- =\sqrt{\varepsilon} \frac{1}{2}(\sigma_x-i\sigma_y)$,
often referred to as spontaneous emission.
For the estimation of $\omega$, it is straightforward to verify that the DHLS condition is satisfied for both Hamiltonians. In this case, for $T\to\infty$ we have
\begin{equation}
    \mathcal{B}_{\mathrm{AC}}=\frac{8B^2}{3\varepsilon}T^3,
    \qquad
    \mathcal{B}_{\mathrm{RF}}=\frac{2B^2(8+3\pi)}{3\pi\varepsilon}T^3.
    \label{eq:DHLS_example_largeT}
\end{equation}
As before, the maximum is taken over all possible protocols described in the previous section.
Unlike in the DHNLS case, the time scaling is now reduced to $T^3$, which nevertheless still surpasses the $T^2$ Heisenberg scaling characteristic of time-independent signals.

Again, a full numerical evaluation of the bound is possible through the SDP method (see SM \cite{supplementary}), and it is plotted in Fig.~\ref{fig:sdp_bounds} for both signals.
In Fig.~\ref{fig:sdp_bounds} one can observe a transient regime after which the QFI crosses over from the short-time $T^4$ scaling, to the asymptotic $T^3$ behavior given by Eq.~\eqref{eq:DHLS_example_largeT}.

\begin{figure}[t]
    \centering
    \includegraphics[width=1\linewidth]{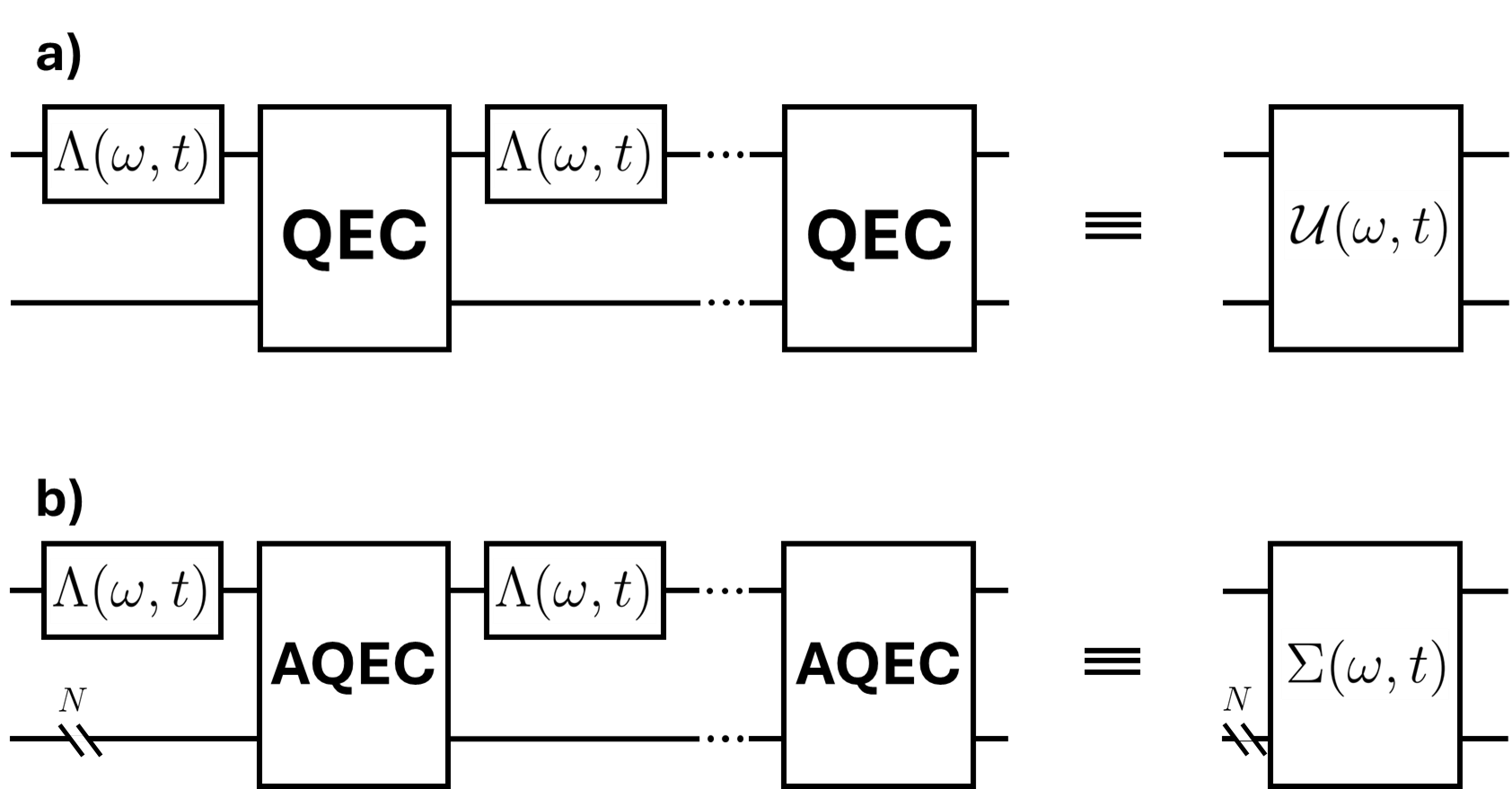}
    \caption{a) Pictorial representation of the QEC scheme to achieve the bound Eq.~\eqref{eq:bound_DHLS_DHNLS} in the DHNLS scenario.
    A QEC step is applied periodically, coupling the probe to a noiseless ancilla, the effective dynamics becomes a unitary $\mathcal{U}(\omega, t)$ as discussed in the main text. 
    b) Pictorial representation of the AQEC scheme to achieve the bound Eq.~\eqref{eq:bound_DHLS_DHNLS} in the DHLS scenario. An AQEC step is applied periodically, coupling the probe to a $N$ noiseless ancillas, the effective dynamics becomes a channel $\Sigma(\omega,t)$, whose effect in logical space is described by Eq.~\eqref{eq:gksl_time_dep_eps_text}.}
    \label{fig:scheme}
\end{figure}

\textit{QEC protocol in the DHNLS regime.}---In the DHNLS regime one can construct an asymptotically optimal metrological strategy that saturates the DHNLS branch of the bound in Eq.~\eqref{eq:bound_DHLS_DHNLS} for large interrogation time $T$, by means of QEC.
The scheme is illustrated in Fig.~\ref{fig:scheme} a): we couple the probe to a noiseless ancilla and apply a periodic QEC step, so that, within an appropriate code space, the effective evolution in logical space is coherent, in close analogy with the time-independent scenario~\cite{Zhou2018,Zhou2021}.

Let $\mathcal{H}_S$ be the Hilbert space of the probe and $\mathcal{H}_A$ the Hilbert space of the ancilla, with $\mathcal{H}_A\simeq \mathcal{H}_S$.
We consider a time-dependent two-dimensional code subspace $\mathcal{C}(t)=\mathrm{span}\bigl\{\ket{c_0(t)},\ket{c_1(t)}\bigr\}\subset \mathcal{H}_S\otimes \mathcal{H}_A,
\braket{c_i(t)}{c_j(t)}=\delta_{ij}$ and we denote by $\sigma_{z,\mathrm{L}}(t)=\ket{c_0(t)}\!\bra{c_0(t)}-\ket{c_1(t)}\!\bra{c_1(t)}, \Pi(t) = \ket{c_0(t)}\!\bra{c_0(t)}+\ket{c_1(t)}\!\bra{c_1(t)}, \Pi_\perp(t) = \id-\Pi(t)$ respectively the logical Pauli-$z$ operator on $\mathcal{C}(t)$, the projection onto $C(t)$ and the projection on its complement.
Every QEC step will consist in the infinitely fast application of $\Pi(\cdot) \Pi+ \mathcal{R} (\Pi_\perp (\cdot) \Pi_\perp)$, where $\mathcal{R}$ is a CPTP map describing the recovery channel.
It is possible to prove (see SM \cite{supplementary}) that, under periodic application of the QEC step, namely every small timestep $dt$~\footnote{Here small is intended with respect to the system energy scales but not small enough to break the Lindblad representation.}, the reduced dynamics within the code takes the effective Hamiltonian form (taking the limit $dt \rightarrow 0$) 
\begin{equation}
\dot{\rho}_{\mathrm{L}}(t)
=
-i\left[
\frac{1}{2}\,\Tr\!\bigl(H_{\perp}(t)\,\sigma_{z,\mathrm{L}}(t)\bigr)\,\sigma_{z,\mathrm{L}}(t),
\,\rho_{\mathrm{L}}(t)
\right],
\label{eq:effective_logical_dyn}
\end{equation}
where $\rho_{\mathrm{L}}(t)$ is the logical two-dimensional state, and $H_{\perp}(t)$ is the projection of $H(\omega,t)$ onto the orthogonal complement of the Lindblad span $\mathcal{S}(t)$ with respect to the Hilbert--Schmidt inner product.

Since the dynamics of Eq.~\eqref{eq:effective_logical_dyn} is coherent, it is possible to apply tools similar to those derived in~\cite{Pang2017}
% \cite{pang_optimal_2017}
(see SM \cite{supplementary}) to obtain the optimally controlled QFI, which, for a given choice of the code space, will read
\begin{equation}
Q(T)
=
\left[
\int_{0}^{T}\Tr\bigl(H'_{\perp}(t)\,\sigma_{z,\mathrm{L}}(t)\bigr)\,dt
\right]^2 .
\label{eq:qfi_code_dependent}
\end{equation}
% Deriving the corresponding optimal controls requires distinguishing two situations (see SM \cite{supplementary}). If the code subspace is time-independent as a subspace, i.e.$
% \mathrm{span}\{\ket{c_0(t)},\ket{c_1(t)}\}
% =
% \mathrm{span}\{\ket{c_0(t')},\ket{c_1(t')}\}
% $ $ \forall\, t,t',
% \label{eq:time_independent_codespace}$
% then one can simply apply continuously a control Hamiltonian.
% \begin{align} \label{eq:control_ham_DHNLS}
%     H_c(t) & \notag = i ( |\partial_t c_0 (t)\rangle \langle c_0(t)| +  |\partial_t c_1(t) \rangle \langle c_1(t)| ) \\ &=  \langle c_1(t)| \partial_t c_0(t) \rangle \sigma_{yL}(t) \,.
% \end{align}
% If instead the code subspace itself changes with time, a constant in-code Hamiltonian component would be removed by the subsequent QEC recovery; consequently, one must implement an additional unitary update after each recovery step to track the time-dependent code, which will mimic the effect of the control Hamiltonian Eq.~\eqref{eq:control_ham_DHNLS}. In particular the unitary control reads:
% \begin{align}
%     &V_C(t+dt)   \notag= \mathcal{T}\exp{-i \int_{t}^{t+dt} H_C(t')dt'  } \\ & \quad \approx  | c_0(t+dt) \rangle \langle c_0(t) |+ | c_1(t+dt) \rangle \langle c_1(t) |\,.
% \end{align}
% Interestingly, the code space is always time independent for two-levels probes.
Deriving the corresponding optimal controls requires distinguishing two situations (see SM \cite{supplementary}).
If the code projector is time-independent, $\Pi(t)=\Pi$, the basis can be tracked by a continuous in-code Hamiltonian
\begin{align} \label{eq:control_ham_DHNLS}
    H_c(t)
    =& \notag
    \sum_{j=0,1}\!\left[f_j(t)-E_j(\omega_c,t)\right]
    \ket{c_j(t)}\!\bra{c_j(t)}
    \\
    &+ i\sum_{j=0,1}
    \ket{\partial_t c_j(t)}\!\bra{c_j(t)} ,
\end{align}
where $E_j(\omega_c,t)=(-1)^j\Tr[H_\perp(\omega_c,t)\sigma_{z,\mathrm{L}}(t)]/2$ and the real functions $f_j(t)$ set only known phases.
If instead the code projector itself changes with time, the tracking is implemented as a unitary update after each recovery step,
\begin{align}
    V_C(t+dt)
    \approx
    \sum_{j=0,1}\ket{c_j(t+dt)}\!\bra{c_j(t)}
    +V_\perp(t,t+dt),
\end{align}
with $V_\perp$ completing the map to a unitary.
For a bare two-level probe without encoding the span of two instantaneous states is the whole probe Hilbert space; in the encoded QEC construction with ancillas, however, the
optimal code projector can genuinely depend on time.

Finally, maximizing \eqref{eq:qfi_code_dependent} over all admissible code choices, one finds (see SM~\cite{supplementary})
% {\footnotesize 
\begingroup
\footnotesize
\setlength{\belowdisplayskip}{4pt}
\setlength{\belowdisplayshortskip}{4pt}
\begin{align}
\max_{\ket{c_0(t)},\,\ket{c_1(t)}} Q(T)
&=
\left[
\int_{0}^{T}
\max_{\ket{c_0(t)},\,\ket{c_1(t)}}
\Tr\!\bigl(H'_{\perp}(t)\,\sigma_{z,\mathrm{L}}(t)\bigr)\,dt
\right]^2
\nonumber\\
&=
4\left[
\int_{0}^{T}
\min_{\kappa(t),\,\eta(t),\,\gamma(t)}
\bigl\|\beta(t)\bigr\|\,dt
\right]^2 ,
\label{eq:qec_saturates_dhnls}
\end{align}
% }
\endgroup
which is valid in the asymptotic long-time regime and saturates the DHNLS branch of the bound in Eq.~\eqref{eq:bound_DHLS_DHNLS}.

\textit{Approximate QEC in the DHLS regime.}---Regarding the opposite regime, the DHLS branch of the bound in Eq.~\eqref{eq:bound_DHLS_DHNLS} can be saturated by means of AQEC~\cite{Zhou2020},
% \emph{approximate} quantum error correction (AQEC)
% ~\cite{zhou2020optimal}, 
with the scheme pictorially represented in Fig. \ref{fig:scheme} b).
As proven in the SM \cite{supplementary}, suitably chosen spin-squeezed probe states can saturate the precision bounds for evolutions of the form
\begin{equation}\label{eq:gksl_time_dep_eps_text}
    \dot{\rho}  = -\frac{i}{2}F(\omega, t) [\sigma_z, \rho]+ \varepsilon(t)( \sigma_z \rho \sigma_z - \rho) \,.
\end{equation}
The goal is therefore to engineer, via periodic AQEC steps acting on the probe and noiseless ancillas every small timestep $dt$, an effective logical dynamics that is equivalent to Eq.~\eqref{eq:gksl_time_dep_eps_text}: a signal along a fixed logical $z$ axis with a time-dependent amplitude, affected by pure dephasing along the same logical axis with a time-dependent rate $\varepsilon_{\mathrm{L}}(t)$.

Let $\mathcal{H}_S$ be the Hilbert space of the probe and $\mathcal{H}_A$ that of a noiseless ancilla with $\mathcal{H}_A\simeq \mathcal{H}_S$. In addition, we introduce a further noiseless two-dimensional ancillary space $\mathcal{H}_O$.
Similarly to the previous section we consider a time-dependent two-dimensional code subspace $ \mathcal{C}(t)=\mathrm{span}\bigl\{\ket{c_0(t)},\ket{c_1(t)}\bigr\}\subset \mathcal{H}_S\otimes \mathcal{H}_A\otimes \mathcal{H}_O, \; \braket{c_i(t)}{c_j(t)}=\delta_{ij}$.
Again, we consider the logical Pauli-$z$ operator on $\mathcal{C}(t)$, the projection onto $C(t)$ and the projection on its complement, defined as in the previous section and similarly the AQEC step will be described by  $\Pi(\cdot) \Pi+ \mathcal{R} (\Pi_\perp (\cdot) \Pi_\perp)$.
However, now we will focus on a specific class of recovery channels~\cite{Zhou2020}, which will be shown to be general enough for our purposes (see SM \cite{supplementary}):
% Namely:
\begin{align} 
    \mathcal{R}_t ( \cdot ) & = \sum_m \notag \left( | c_0(t) \rangle \langle R_m(t), 0| + | c_1(t) \rangle \langle Q_m (t), 1 | \right) ( \cdot)\\& \left(|R_m(t),0 \rangle \langle c_0(t)| + |Q_m(t) , 1 \rangle \langle c_1(t)| \right) \,, 
\end{align}
where $\{ | Q_m \rangle \},\{ | R_m \rangle \} $ are two orthonormal bases of $\mathcal{H}_S \otimes \mathcal{H}_A$.
By applying the QEC step periodically every small timestep $dt$ one obtains an effective logical GKSL equation of the form (taking the limit for $dt \rightarrow 0$)
\begin{align}
 \notag \frac{d\rho_{\mathrm{L}}(t)}{dt}&
=
-i\left[
\frac{1}{2}\,\Tr\!\bigl(H(\omega,t)\,\sigma_{z,\mathrm{L}}(t)\bigr)\,\sigma_{z,\mathrm{L}}(t),
\,\rho_{\mathrm{L}}(t)
\right]
\\ &+\varepsilon_{\mathrm{L}}(t)\Bigl(\sigma_{z,\mathrm{L}}(t)\,\rho_{\mathrm{L}}(t)\,\sigma_{z,\mathrm{L}}(t)-\rho_{\mathrm{L}}(t)\Bigr),
\label{eq:aqec_time_dep_axis}
\end{align}
where $\varepsilon_L(t)\ge 0$ is the effective logical coupling whose explicit expression is given in the SM \cite{supplementary}.
Eq.~\eqref{eq:aqec_time_dep_axis} can be viewed locally as pure dephasing with a direction (and strength) that varies with time. 

To constrain the effective dynamics to be in the form of Eq.~\eqref{eq:gksl_time_dep_eps_text}, we further use a control step that \emph{aligns} the instantaneous logical axis to a fixed one.
Concretely, after each AQEC recovery we apply an additional control given by:
\begin{align}
V_\mathrm{C}(t+dt) =& \notag | c_0(t+dt) \rangle \langle c_0(t) | \\&+ | c_1(t+dt) \rangle \langle c_1(t) |,
\end{align}
steering the dynamics to follow the evolution of $|c_{0(1)}(t) \rangle$. We assume the implementation of $V_C$ to employ negligible time and $dt$ to be the small time step elapsed between two QEC steps.

The dephasing direction becomes time-independent in the logical frame, indeed, the resulting effective dynamics is
\begin{align}
\dot{\rho}_{\mathrm{L}}(t)&
=  \notag
-i\left[
\frac{1}{2}\,\Tr\!\bigl(H(\omega,t)\,\sigma_{z,\mathrm{L}}(t)\bigr)\,\sigma_{z,\mathrm{L}}(T),
\,\rho_{\mathrm{L}}(t)
\right]
\\ & +\varepsilon_{\mathrm{L}}(t)\Bigl(\sigma_{z,\mathrm{L}}(T)\,\rho_{\mathrm{L}}(t)\,\sigma_{z,\mathrm{L}}(T)-\rho_{\mathrm{L}}(t)\Bigr),
\label{eq:PD_aqec}
\end{align}
i.e.\ a signal and pure dephasing along a \emph{fixed} logical operator $\sigma_{z,\mathrm{L}}(T)$ (the dependence on $T$ is set by the final alignment choice), with a time-dependent dephasing rate $\varepsilon_{\mathrm{L}}(t)$.

The dynamics described by Eq.~\eqref{eq:PD_aqec} is therefore of the form of a dephasing in a constant direction.
Therefore, as proved in the SM \cite{supplementary}, employing a one-axis-twisted spin-squeezed state (OATSSS) in a Ramsey-type scheme the QFI reads:

% \begin{equation}
%         Q(T)= \int_{0}^{T}
% \frac{\Tr\!\bigl(H(\omega,t)'\,\sigma_{z,\mathrm{L}}(t)\bigr)^{2}}{\varepsilon_{\mathrm{L}}(t)}\,dt
% \end{equation}
% \FA{\begin{equation}
%         Q(T)= \int_{0}^{T}
% \frac{\Tr\!\bigl(H(\omega,t)'\,\sigma_{z,\mathrm{L}}(t)\bigr)^{2}}{4\varepsilon_{\mathrm{L}}(t)}\,dt
% \end{equation}}
\begin{equation}
        Q(T)= \int_{0}^{T}
\frac{\Tr\!\bigl(H'(\omega,t)\,\sigma_{z,\mathrm{L}}(t)\bigr)^{2}}{4\varepsilon_{\mathrm{L}}(t)}\,dt
\end{equation}
Interestingly, as proved by Zhou et al. in~\cite{Zhou2020}, it is possible to reduce to a particular shape of the code space.
Namely, $|c_0 / c_1 \rangle = \sum_{ij} A_{0/1 ij} | i \rangle | j \rangle | 0/1 \rangle\,,$ such that $A_0, A_1 \in \mathbb{C}^{d \times d}$, $A_{0,ij} = \sqrt{1-\varepsilon^2}+ \varepsilon D_{ij}$ and $A_{1,ij} = \sqrt{1-\varepsilon^2}- \varepsilon D_{ij}$ satisfy $\Tr[A_0^\dag A_0] =\Tr[A_1^\dag A_1] =1 $ and $\Tr[C^\dag D] = 0$.

Finally, we can exploit the time-independent result of Zhou et al.~\cite{Zhou2020} and write:
\begin{align}
&\notag\max_{\ket{c_0(t)},\,\ket{c_1(t)}}  \lim_{\varepsilon \rightarrow 0} Q(T)
\\&=
\int_{0}^{T}
\max_{\ket{c_0(t)},\,\ket{c_1(t)}}
%  \lim_{\varepsilon \rightarrow 0 } \frac{\Tr\!\bigl(H(\omega,t)'\,\sigma_{z,\mathrm{L}}(t)\bigr)^{2}}{\varepsilon_{\mathrm{L}}(t)}\,dt
	% \FA{ \lim_{\varepsilon \rightarrow 0 } \frac{\Tr\!\bigl(H(\omega,t)'\,\sigma_{z,\mathrm{L}}(t)\bigr)^{2}}{4\varepsilon_{\mathrm{L}}(t)}\,dt}
	 \lim_{\varepsilon \rightarrow 0 } \frac{\Tr\!\bigl(H'(\omega,t)\,\sigma_{z,\mathrm{L}}(t)\bigr)^{2}}{4\varepsilon_{\mathrm{L}}(t)}\,dt
\nonumber\\
&=
4 \int_{0}^{T}
\min_{\substack{\kappa(t),\,\eta(t),\,\gamma(t)\\ \beta(t)=0}}
\bigl\|\alpha(t)\bigr\|\,dt,
\label{eq:aqec_saturates_dhls}
\end{align}
showing that the DHLS branch of the bound in Eq.~\eqref{eq:bound_DHLS_DHNLS} is achieved also in the DHLS regime.

\emph{Conclusion}.---In summary, we have established the ultimate precision bounds for estimating parameters encoded in time-dependent signals subject to Markovian noise.
By generalizing channel extension methods to dynamically varying environments, we provide a unified framework whose fundamental limits can be efficiently evaluated at all times via semidefinite programming.
Crucially, we demonstrate a universal scaling law (under mild regularity conditions): if a noiseless time-dependent evolution yields a quantum Fisher information (QFI) scaling as $T^{2k}$, the presence of noise satisfying the HNLS condition preserves this $T^{2k}$ scaling, whereas noise satisfying the HLS condition fundamentally restricts the scaling to $T^{2k-1}$.
This result generalizes the well-known time-independent boundaries of the Heisenberg and standard quantum limits to the dynamically driven regime.

To prove the physical achievability of these bounds, we introduced generalized error-correction protocols.
For systems in the HNLS regime, continuous QEC, supplemented by adaptive control tracking the time-dependent code space, yields an effective coherent evolution that asymptotically saturates the ultimate bound.
In the HLS regime, we proved that the corresponding bounds can be saturated using AQEC combined with spin-squeezed states in a generalized Ramsey sequence. 

Our findings confirm that the metrological advantages generated by time-dependent modulation survive in open quantum systems.
Translating these theoretical limits into practical devices, however, requires the development of more realistic QEC protocols~\cite{Zhou2024c,Sahu2026,Chen2026n}.
Further research is also needed to extend this framework to non-Markovian noise~\cite{Mann2025,Kurdzialek2025a}.
% the finite-shot regime, where asymptotic statistical bounds may not immediately apply, and 
Finally, we have assumed, as is customary in this field, that time-dependent control Hamiltonians do not fundamentally alter the noise model.
In practice, microscopically derived noise models can depend on the applied control Hamiltonians, and the metrological implications of this observation have only recently begun to be explored~\cite{Gorecki2026}.

\textit{Acknowledgments.}---FA would like to thank \mbox{R.~{Demkowicz-Dobrza{\'n}ski}} and \mbox{M.~G.~Genoni} for many useful discussions.

\bibliography{biblio_FA}

\end{document}